\documentstyle[psfig,graphicx]{mn}

\def\degr{\hbox{$^\circ$}}
\def\>{$>$}
\def\<{$<$}
\def\simlt{\lower.5ex\hbox{$\; \buildrel < \over \sim \;$}}
\def\simgt{\lower.5ex\hbox{$\; \buildrel > \over \sim \;$}}

\newif\ifAMStwofonts

\def\simlt{\lower.5ex\hbox{$\; \buildrel < \over \sim \;$}}
\def\simgt{\lower.5ex\hbox{$\; \buildrel > \over \sim \;$}}

\ifoldfss
  \ifCUPmtlplainloaded \else
    \NewTextAlphabet{textbfit} {cmbxti10} {}
    \NewTextAlphabet{textbfss} {cmssbx10} {}
    \NewMathAlphabet{mathbfit} {cmbxti10} {} 
    \NewMathAlphabet{mathbfss} {cmssbx10} {} 
  \fi
  \ifAMStwofonts
    \ifCUPmtlplainloaded \else
      \NewSymbolFont{upmath} {eurm10}
      \NewSymbolFont{AMSa} {msam10}
      \NewMathSymbol{\upi}     {0}{upmath}{19}
      \NewMathSymbol{\umu}     {0}{upmath}{16}
      \NewMathSymbol{\upartial}{0}{upmath}{40}
      \NewMathSymbol{\leqslant}{3}{AMSa}{36}
      \NewMathSymbol{\geqslant}{3}{AMSa}{3E}

      \let\leq=\leqslant 
      \let\geq=\geqslant 
    \fi
  \fi
\fi 

\ifnfssone
  \newmathalphabet{\mathit}
  \addtoversion{normal}{\mathit}{cmr}{m}{it}
  \addtoversion{bold}{\mathit}{cmr}{bx}{it}
  \newmathalphabet{\mathbfit} 
  \addtoversion{normal}{\mathbfit}{cmr}{bx}{it}
  \addtoversion{bold}{\mathbfit}{cmr}{bx}{it}
  \newmathalphabet{\mathbfss} 
  \addtoversion{normal}{\mathbfss}{cmss}{bx}{n}
  \addtoversion{bold}{\mathbfss}{cmss}{bx}{n}
  \ifAMStwofonts
    \ifCUPmtlplainloaded \else
      \UseAMStwoboldmath
      \makeatletter
      \new@mathgroup\upmath@group
      \define@mathgroup\mv@normal\upmath@group{eur}{m}{n}
      \define@mathgroup\mv@bold\upmath@group{eur}{b}{n}
      \edef\UPM{\hexnumber\upmath@group}
      \new@mathgroup\amsa@group
      \define@mathgroup\mv@normal\amsa@group{msa}{m}{n}
      \define@mathgroup\mv@bold\amsa@group{msa}{m}{n}
      \edef\AMSa{\hexnumber\amsa@group}
      \makeatother
      \mathchardef\upi="0\UPM19
      \mathchardef\umu="0\UPM16
      \mathchardef\upartial="0\UPM40
      \mathchardef\leqslant="3\AMSa36
      \mathchardef\geqslant="3\AMSa3E

      \let\leq=\leqslant 
      \let\geq=\geqslant 
    \fi
  \fi
\fi 

\ifnfsstwo
  \DeclareMathAlphabet{\mathbfit}{OT1}{cmr}{bx}{it}
  \SetMathAlphabet\mathbfit{bold}{OT1}{cmr}{bx}{it}
  \DeclareMathAlphabet{\mathbfss}{OT1}{cmss}{bx}{n}
  \SetMathAlphabet\mathbfss{bold}{OT1}{cmss}{bx}{n}
  \ifAMStwofonts
    \ifCUPmtlplainloaded \else
      \DeclareSymbolFont{UPM}{U}{eur}{m}{n}
      \SetSymbolFont{UPM}{bold}{U}{eur}{b}{n}
      \DeclareSymbolFont{AMSa}{U}{msa}{m}{n}
      \DeclareMathSymbol{\upi}{0}{UPM}{"19}
      \DeclareMathSymbol{\umu}{0}{UPM}{"16}
      \DeclareMathSymbol{\upartial}{0}{UPM}{"40}
      \DeclareMathSymbol{\leqslant}{3}{AMSa}{"36}
      \DeclareMathSymbol{\geqslant}{3}{AMSa}{"3E}

      \let\leq=\leqslant 
      \let\geq=\geqslant 
    \fi
  \fi
\fi 

\ifCUPmtlplainloaded \else
  \ifAMStwofonts \else 
    \def\upi{\pi}
    \def\umu{\mu}
    \def\upartial{\partial}
  \fi
\fi

\title[Flickering studies of X1822-371]
		{Multicolor flickering studies of X1822-371}
\author[R. Baptista et~al.]
       {Raymundo Baptista$^1$, A. Bortoletto$^1$ and E. T. Harlaftis$^2$ \\
       $^1$ Departamento de F\'\i sica, Universidade Federal de Santa Catarina,
       Campus Trindade, 88040-900, Florian\'opolis - SC, Brazil, \\
       ~ email: bap@fsc.ufsc.br, alex@fsc.ufsc.br \\
       $^2$ Institute of Astronomy and Astrophysics, National Observatory 
       of Athens, P.O. Box 20048, Thession -- 110 48, Athens, Greece, \\ 
       ~ email: ehh@astro.noa.gr }
\date{Accepted 2002 May 8; Received ????; in original form 2002 February 19}

\pagerange{1--8}
\pubyear{2002}

\begin{document}

\label{firstpage}

\maketitle

\begin{abstract}

We report on the analysis of high-speed multicolor photometry of the
eclipsing X-ray binary X1822-371. 
We used new eclipse timings to derive a revised optical ephemeris. 
A quadratic fit to the eclipse timings is not statistically
significant but suggests that the orbital period is increasing on a
timescale of $P_0/|\hbox{\.{P}}|= (4.2 \pm 1.4) \times 10^6$~yr.
We find no systematic delay or advance of the optical timings with 
respect to the X-ray timings.
Average UBVRI light curves show the deep eclipse of the disc by the 
secondary star superimposed on the broader and shallower occultation
of the inner disc regions by the outer disc (dip), and an orbital hump
centred at phase +0.25 which is mostly seen in the U and B bands. 
There is no evidence of a secondary eclipse at phase +0.5 or of an 
ellipsoidal modulation of the secondary star light.
The starting phase of the dip occurs {\em earlier} for shorter 
wavelengths, while the egress occurs at the same phase in all bands. 
This suggests that the thickening of the outer, occulting disc rim is 
gradual with azimuth at ingress but decreases sharply at egress.
We fit synthetic photometry to the extracted colors of the inner and
outer disc regions to estimate their effective temperatures. We find
$T_{\rm eff}= (9\pm 5)\times 10^7\;K$ and $T_{\rm eff}= (6\pm 2)
\times 10^4\;K$, respectively, for the inner and outer disc regions.
The orbital dependency of the flickering activity is derived from the 
mean scatter of the individual light curves with respect to the average
UBVRI light curves. The flickering curves show a broad eclipse at the
dipping phases, the depth of which decreases with increasing wavelength. 
The blue, eclipsed flickering component is associated with the inner disc
regions and can be fitted by a blackbody spectrum of $T_{\rm eff}= (2.1
\pm 0.8) \times 10^8\;K$, whereas the uneclipsed flickering component 
probably arises from the outermost disc regions and is well described by 
a blackbody of $T_{\rm eff}= (9.6\pm 0.7)\times 10^3\;K$.
\end{abstract}


\section{Introduction}

Low-mass X-ray binaries (LMXB) are close binary systems in which a
late-type star (the secondary) overfills its Roche lobe and transfers
matter to a companion neutron star or black hole (the primary) via an 
accretion disc. As a consequence of the deep gravitational potential well 
of the compact primary star, most of the accretion luminosity is radiated
as X-rays (Lewin, van Paradijs \& van den Heuvel 1995).

The light curves of interacting binaries, active galactic nuclei and
of some T Tauri stars show intrinsic brightness fluctuation of $0.01-1$ 
magnitudes on timescales ranging from seconds to dozens of minutes, 
termed ``flickering''.
Although flickering activity is considered a fundamental signature of 
accretion, it is also the least understood aspect of the accretion 
processes (e.g., Warner 1988, 1995).

X1822-371 is the brightest eclipsing LMXB currently known. Its X-ray
light curve shows a 5.57~hr modulation which is caused by the occultation
of the disc by the companion star (eclipse) and by a bulge (dip) 
associated with the impact region of the gas stream from the secondary
star with the disc rim (White \& Holt 1982; Hellier \& Mason 1989).
In eclipsing LMXBs, the vertically-extended disc can permanently hide
the compact object from our view, leading to a much weaker X-ray source
relative to its optical brightness (p.ex., $L_x/L_{opt}= 20$ in 
X1822-371). Systems like X1822-371 are termed Accretion Disc Corona 
(ADC) sources because the X-rays from these systems are visible through
scattering from a disc corona or wind. Their optical properties can be
understood in terms of reprocessing of X-rays in the accretion disc 
and the companion star.

The empirical model of components from the inner and outer disc by
Hellier \& Mason (1989) yields an estimate of the size of the ADC 
in X1822-371 of $R_{ADC} \sim 0.47\;R_\odot$, an inclination in the 
range $i= 82\degr - 87\degr$, and a disc half-thickness of 
$h_d= 0.08 - 0.22\;R_\odot$. 
Harlaftis et~al. (1997) found a lower limit to the radial velocity
of the secondary star of $K_s> 225\;km\;s^{-1}$, suggesting a spectral 
type M. Cowley et~al. (1982) derived a binary mass ratio
in the range $3 < q(=M_x/M_s) < 5$, resulting in a secondary star mass
of $0.13-0.33\;M_\odot$, and suggested that the secondary star is
undermassive for its radius and luminosity.
The distance to X1822-371 is rather uncertain.
A lower limit of $d=600\;pc$ was found by Mason et~al. (1980) from 
the reddening of the colors. Mason \& Cordova (1982a) fitted 
ultraviolet, optical and infrared light curves to find a distance
of $d=1-5\;Kpc$, with most probable value between $2-3\;Kpc$.

Currently, there is only a handful of isolated light curves of X1822-371
separated by years and mostly measured at poor time resolutions ($\simgt
5$~min), which prevents a more comprehensive study of this binary in the
optical.

In this paper we report on the analysis of high speed optical photometry 
of X1822-371. The observations and data reduction procedures are 
described in section~\ref{data}. We obtain a revised optical ephemeris
and estimate of the orbital period change in section~\ref{efem}. 
In section~\ref{curvas} we present and discuss mean orbital light curves 
and derive the orbital dependency of the optical flickering activity
as a function of wavelength.
In section~\ref{fluxos} we use average UBVRI fluxes at selected phases 
to separate the emission from the different light sources in the system. 
The dependency of the optical flickering with frequency is investigated
in section~\ref{fft}.
A summary of the main results is presented in section~\ref{fim}.

\section{Observations and data reduction} \label{data}

X1822-371 was observed with the 1.6-m and 0.6-m telescopes of the
Laborat\'orio Nacional de Astrof\'\i sica (LNA/CNPq), in southern Brazil,
between 1996 and 2001. We collected a total of nine light curves
covering seven eclipses. The 1996 data were obtained with a photoelectric
photometer (FOTRAP), and the remaining data sets were obtained with
a CCD camera. Table~\ref{jornal} contains a summary of the observations.
Column $\Delta t$ lists either the time resolution (FOTRAP runs) or
the integration time (CCD runs). The last column describes the quality of
the observing night. 
%
\begin{table*}
 \centering
 \begin{minipage}{120mm}
  \caption{Journal of the observations.}
  \label{jornal}
\begin{tabular}{@{}lrcccccc@{}}
~~~~Date & Start & No. of & $\Delta t$ & Phase range & Instrument &
Telescope & Quality
\footnote{ A= photometric (main comparison stable), B= good (some sky
variations), C= poor (large variations and/or clouds).} \\ [-0.5ex]
& (UT) & exposures & (s) \\ [1ex]
15 Jun 1996 &  4:36 & ~665 & 16 & $-0.59,-0.06$ & FOTRAP & 1.6~m & A \\
            & 23:51 & ~467 & 16 & $-0.13,+0.24$ & FOTRAP & 1.6~m & A \\
16 Jun 1996 &  5:12 & ~727 & 16 & $-0.18,+0.40$ & FOTRAP & 1.6~m & A \\
17 Jun 1996 & 23:38 & 1003 & 16 & $-0.56,+0.24$ & FOTRAP & 1.6~m & B \\
18 Jun 1996 &  4:32 & ~900 & 16 & $-0.68,+0.04$ & FOTRAP & 1.6~m & B \\
[0.5ex]
15 Jul 1999 &  4:40 & ~172 & 20 & $-0.21,+0.06$ & CCD & 1.6~m & C \\
30 Jul 2000 & 22:18 & ~117 & 30 & $-0.11,+0.12$ & CCD & 0.6~m & C \\
25 Aug 2000 & 22:07 & ~144 & 35 & $-0.12,+0.17$ & CCD & 0.6~m & B \\ 
29 Jun 2001 &  2:35 & ~670 & 20 & $-0.67,+0.15$ & CCD & 1.6~m & C \\ [-3ex]
\end{tabular}
\end{minipage}
\end{table*}

FOTRAP is a one-channel photometer with a high-speed (20~Hz) rotating 
filter wheel that allows quasi-simultaneous measurements in 6 bands. 
Its photometric system closely matches the Johnson-Cousins UBVRI system 
as defined by Bessell (1990). The remaining position in the filter wheel
is used for observations in white light (W). 
The observations were performed with a small 9.1~arcsec diaphragm to 
avoid contamination by the light from a nearby companion star. 
Sky measurements were taken at intervals of $20-30$~min at a position
40~arcsec to the east of the variable, except during eclipse. The sky
measurements were fitted by a cubic spline function and then subtracted 
from the raw data. A close comparison star was also observed at intervals
of $20-30$~min to check for the presence of clouds and sky transparency
variations. 
From these observations, we confirmed that the FOTRAP runs were 
performed  under good sky conditions.
Coefficients of extinction and of transformation to the UBVRI standard
system for each night were derived from observations of E-regions stars
of Graham (1982) and blue spectrophotometric standards of Stone \& 
Baldwin (1983).
The reader is referred to Jablonski et al. (1994) for a detailed study
of the reliability of the transformations from FOTRAP's natural to the
standard UBVRI system, in particular for objects of peculiar blue 
spectra such as cataclysmic variables and LMXBs.
We used the relations of Lamla (1981) to transform UBVRI magnitudes 
to flux density units. The absolute calibration of the
observations is accurate to better than 10 per cent.

Time-series of differential photometry of X1822-371 in the B-band
were obtained with an EEV CCD camera ($385 \times 578$ pixels, 0.58
arcsec/pixel) in 1999, 2000 and 2001. The observations consisted of
20-35\,s exposures with a negligible dead time between exposures.
The CCD data were reduced using standard {\sc IRAF}\footnote{ 
IRAF is distributed by National Optical Astronomy 
Observatories, which is operated by the Association of Universities 
for Research in Astronomy, Inc., under contract with the National 
Science Foundation. } procedures and included bias and flat-field 
corrections and cosmic rays removal. 
Time-series were constructed by computing the difference
of magnitude of the variable star and of a set of comparison stars 
with respect to a reference star in the field. 
The data were transformed from magnitudes to a flux scale assuming
a unity flux for the reference star.
Most of the CCD runs cover a small phase range around eclipse and were 
included in the analysis only for the purpose of determining a revised 
ephemeris.

\begin{figure}
\includegraphics[bb=1cm 1.5cm 15cm 25cm,scale=0.47]{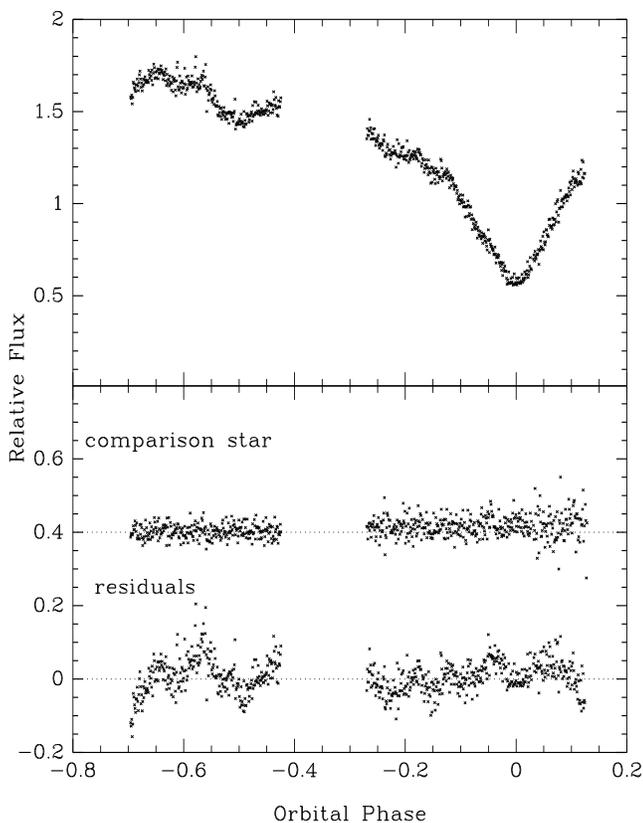}
 \caption{ Top: The B-band light curve of X1822-371 on June 2001, phased 
  according to the linear ephemeris of eq.~\ref{eq:efem}. Bottom: the light
  curve of a comparison star of similar brightness and the residuals with
  respect to a smooth spline function fitted to the light curve of the 
  variable star. The curve of the comparison star was shifted upwards by 
  0.4 units for visualization purposes. }
\label{fig1}
\end{figure}

Figure~\ref{fig1} shows the B-band light curve of X1822-371 on June 2001.
Fluxes are relative to the reference star. The run was interrupted around 
phase $-0.45$ for an observation of a different target star.
The lower panel shows the light curve of a comparison star of similar
brightness and the residuals with respect to a smooth spline function 
fitted to the light curve of the variable star. The scatter around the
mean in the residuals light curve is perceptibly larger than that of 
the comparison star light curve and is caused by flickering. The amplitude
of the flickering around phase $-0.5$ is larger than at mid-eclipse.
%
%

\section{Results and discussion} \label{results}

\subsection{Revised optical ephemeris} \label{efem}

Eclipse timing were determined by fitting parabolic and cubic 
functions to the eclipse shape to estimate times of minimum light,
and by using the bissected chord method to estimate 
times of mid-eclipse (e.g., Baptista, Jablonski \& Steiner 1989).
The final timing is the median of the three measurements and the
corresponding uncertainty is derived from the 
median of the absolute deviations with respect to the median.
The eclipse timings measured from the UBVRI and W-band light curves
are consistent with each other at the 1-$\sigma$ level.
The barycentric correction and the difference between universal times 
(UT) and dynamical ephemeris time scales are smaller than the 
uncertainties in the measured timings and were neglected.
The new heliocentric eclipse timings (HJD) are listed in Table~\ref{tab2}
with corresponding eclipse cycle number (E) and uncertainties (quoted in
parenthesis). These timings were combined with the optical timings
of Hellier \& Mason (1989) to obtain a revised optical ephemeris for
X1822-371. The first optical timing of Hellier \& Mason (1980) is very
uncertain (see Mason et~al. 1980) and was not included in the analysis. 
We found an error in one of the timings of Mason et~al. (1980), possibly
caused by an improper heliocentric correction. The revised epoch of this
timing is HJD 2\,444\,044.861(6)~d ($E=-6766$).
%
\begin{table}
 \centering
 \begin{minipage}{60mm}
  \caption{New eclipse timings of X1822-371.}
  \label{tab2}
\begin{tabular}{@{}clc@{}}
E & ~~~~HJD & (O$-$C)\footnote{with respect to the ephemeris of 
eq.~\ref{eq:efem}} \\ [-0.5ex]
(cycles)& (2\,400\,000 +) & (0.001 d) \\ [1ex]
19970 & 50250.5308(8)  & $-1.51$ \\ 
19971 & 50250.7626(8)  & $-1.81$ \\ 
19979 & 50252.6196(8)  & $-1.69$ \\
24809 & 51373.7094(8)  &  +0.26 \\
26458 & 51756.4577(12) &  +0.35 \\ 
26570 & 51782.4542(12) &  +0.63 \\
27894 & 52089.7692(8)  &  +2.92 \\ [-3ex]
\end{tabular}
\end{minipage}
\end{table}

\begin{figure}
\includegraphics[bb=1.5cm 3cm 20.5cm 24cm,angle=-90,scale=0.38]{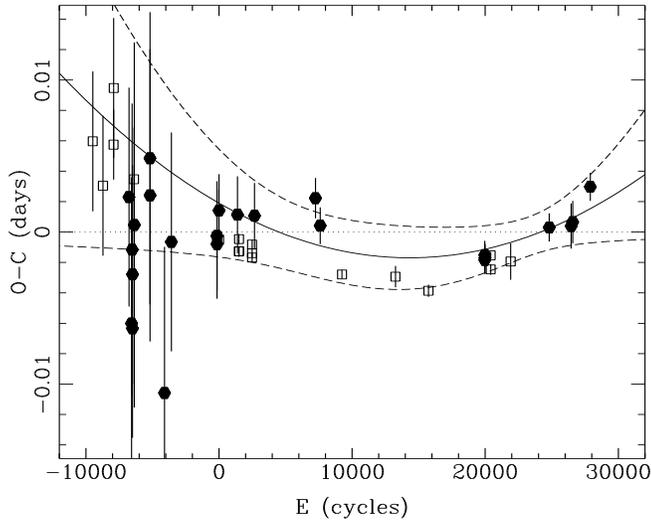}
 \caption{ The (O$-$C) diagram with respect to the linear ephemeris of
 eq.(\ref{eq:efem}). The optical timings are shown as filled circles and
 the X-ray timings are indicated by open squares. Vertical bars represent 
 the uncertainties in the timings. The best-fit quadratic ephemeris is
 displayed as a solid line. The dashed lines show the uncertainties in 
 the quadratic ephemeris at the 3-$\sigma$ level, the covariance between 
 the parameters taken into account. } 
\label{fig2}
\end{figure}

The best-fit least-squares linear ephemeris is,
\begin{equation} 
{\rm T_{mid} = HJD}\;2\,445\,615.310\:(1)\; + 0.232\,109\,28\:(5)
\:{\rm E} \; ,
\label{eq:efem}
\end{equation}
with a dispersion of $\sigma= 0.0034$~d and a reduced $\chi^2$ of 1.27 
for 23 degrees of freedom. The (O$-$C) values with respect to
eq.(\ref{eq:efem}) are listed in Table~\ref{tab2} and plotted in
Fig.~\ref{fig2}. The X-ray eclipse timings of Parmar et~al. (2000)
are also plotted for comparison.
%
%

The best-fit least-squares quadratic ephemeris is,
\begin{eqnarray*} 
{\rm T_{mid}} & = & {\rm HJD}\;2\,445\,615.312\:(1)\; + 0.232\,1088\:(2) 
\:{\rm E} \\
			  &   & + (1.8 \pm 0.6)\times 10^{-11} \:{\rm E}^2 \; ,
\end{eqnarray*}
with resulting $\sigma=0.0057$~d and $\chi^2_{red} = 1.00$.
The quadratic fit results simultaneously in a slightly lower 
$\chi^2_{red}$ and in a significantly larger $\sigma$.
This apparent contradition can be explained by noting that $\sigma$ 
yields a simple mean of the deviations between the fit and the data, 
whereas the $\chi^2$ weights these deviations by the inverse of the 
square of the uncertainty at each data point. This implies that the
quite uncertain earlier timings are essencially neglected in the 
least-squares quadratic fit, but dominate the computation of $\sigma$.

The quadratic ephemeris is not statistically significant (because the
resulting $\sigma$ is even larger than that of the linear fit), but 
suggests a positive variation of the orbital period with a characteristic
timescale of $P_0/|\hbox{\.{P}}|= (4.2 \pm 1.4) \times 10^6$~yr. 
This is consistent with the analysis of the X-ray timings, which 
indicates that the orbital period of X1822-371 increases in a timescale 
of $P_0/|\hbox{\.{P}}|= 3.1 \times 10^6$~yr (Hellier \& Smale 1994).

White et~al. (1981) observed that the optical eclipse timings were
{\em earlier} than the X-ray timings by 13 minutes. However, this 
discrepancy is comparable to the (quite large) uncertainties in their
eclipse timings and in the ephemeris that they adopted. 
At the low time resolution of their data, it is also possible that the 
measured timings are biased towards the mid-dip time, which occurs 
before mid-eclipse time. On the other hand, Hellier \& Mason (1989)
found optical eclipse timings which are {\em late} with respect to 
the X-ray eclipse timings by $\simeq 3$~min and suggested that this 
should be the result of the optical light centre being ahead from the 
line of the star centres.

The (O-C) diagram of Fig.~\ref{fig2} shows that, despite the fact that
the offset between the optical and x-ray timings at the proper epochs
(i.e., around cycles $-8000$ and 2000) are consistent with the reports of,
respectively, White et~al. (1981) and Hellier \& Mason (1989), 
we found no evidence of a systematic delay or advance of the optical 
timings with respect to the X-ray timings.
The initial eclipse times predicted by the optical and X-ray ephemerides
are consistent with each other at the 2-$\sigma$ level.
The larger dispersion of the optical timings in comparison to the
X-ray timings can be understood as a consequence of the fact
that the optical light arises from an extended region of the
accretion disc the light center of which varies in position in response
to time-dependent asymmetries in the surface brightness distribution 
of the outer disc, while the X-ray light is produced in the compact
inner disc regions leading to more stable eclipse timings.
The larger dispersion of the optical timings (and the possible
bias in the earlier timings) is also responsible for the lower 
statistical significance of the quadratic term in the
ephemeris in comparison with the X-rays results.

\subsection{Average light curves} \label{curvas}

We adopted the following convention regarding the phases: conjunction
occurs at phase zero, the phases are negative before conjunction and
positive afterwards. 

\begin{figure*}
\includegraphics[bb=2cm 3cm 20.5cm 24cm,angle=-90,scale=0.65]{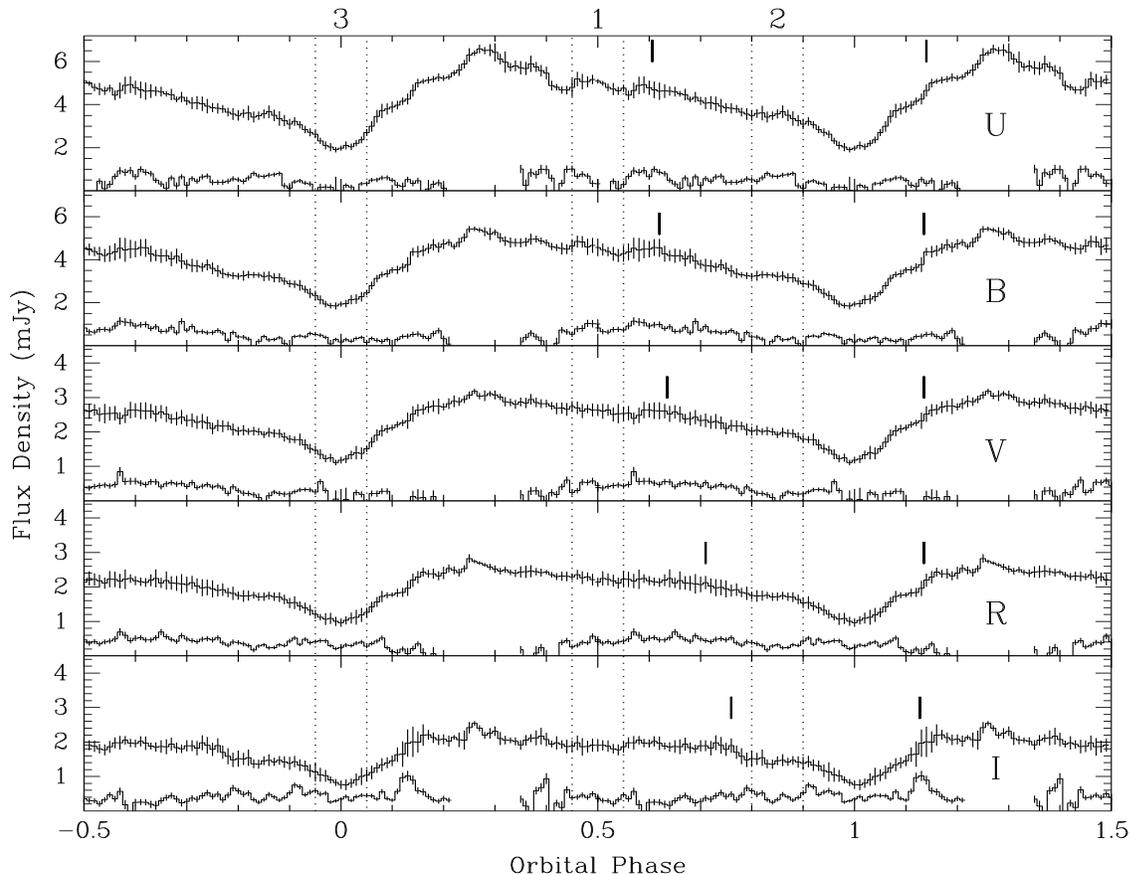}
 \caption{ Average orbital light curves and flickering curves in UBVRI 
  at a resolution of 0.01~cycle. The data are repeated in phase for the
  purpose of visualization. The flickering curves were multiplied by a
  factor of 2 for better visualization. The begin/end of the dip are 
  marked by vertical ticks in each panel. Vertical dotted lines show the
  three regions selected to extract average fluxes (section ~\ref{fluxos}). } 
\label{fig3}
\end{figure*}

In order to increase the signal-to-noise ratio and to reduce the 
influence of flickering in the light curves, we combined the individual
FOTRAP light curves to produce average UBVRI orbital light curves.
For each band, the data is divided into phase bins of width
0.01~cycle, the median flux is computed for each bin and the 
corresponding uncertainty is derived from the median of the 
absolute deviations with respect to the median.
For each average light curve a curve of the root mean square deviations
with respect to the median, $\sigma_{tot}$, is also produced.
These curves have major contributions from the photon count statistics
(which is largely dominant in comparison to the sky scintillation) and 
from the intrinsic flickering activity. The orbital curve of the 
flickering component can be obtained from the relation,
\begin{equation}
\sigma_{flick}(\phi)= \sqrt{\sigma_{tot}^{2}(\phi) -
\sigma_{poisson}^{2}(\phi)} \;\;\;,
\end{equation}
where we adopt $\sigma_{flick}(\phi) = 0$ when $\sigma_{poisson}(\phi) 
\geq \sigma_{tot}(\phi)$ and $\sigma_{poisson}(\phi)$ is determined
from the photon count statistics at the orbital phase $\phi$. 
The flickering component was also set to zero at those phase bins
which consisted of data from less than three light curves.
The computed B-band flickering curve is consistent at the 1-$\sigma$ 
level with the flickering observed in the B-band CCD light curve of June 
2001 (Fig.~\ref{fig1}), both in amplitude and in orbital dependency. 
This give us confidence that the UBVRI flickering curves derived from 
the data collected with the one-channel FOTRAP photometer are not 
affected by an improper correction of fast ($\leq 20$~min), low 
amplitude sky transparency variations. 
The UBVRI average light curves and flickering curves are shown in 
Fig.~\ref{fig3}. 
%
%

The orbital light curves show the deeper eclipse of the disc by the
secondary star between phases $-0.1$ and +0.1 superimposed on a more
extended and shallower eclipse, interpreted as the result of the
occultation of the inner disc regions by a vertically-extended disc
rim (dip), which occurs between phases $-0.4$ and +0.15. The orbital
light curves also display an orbital hump with maximum at phase +0.25,
the amplitude of which decreases for increasing wavelengths.

Mason et~al. (1980) suggest the existence of a shallow secondary
minimum ($\Delta m=0.15-0.2$ mag) centred at orbital phase +0.5
presumably caused by ellipsoidal modulation of the light from the
secondary star, with a possible eclipse of this star by the 
geometrically-thick accretion disc. 
We found no evidence of a secondary eclipse 
(except, perhaps, in the U-band) nor there is evidence of an ellipsoidal 
modulation of the secondary star light.
This is more conspicuous at the longer wavelengths, where the 
dip starts at latter phases leading to a cleaner view of the star's
behaviour around phase +0.5.
This is in contrast with the expected behaviour for an eclipse of
the secondary star by the disc, since the occultation of the red
secondary star should result in progressively deeper eclipses at 
longer wavelengths.
These results indicate that the secondary star in X1822-371 does not
contributes significantly to the optical flux, even in the I-band.
Hence, we associate the decrease in flux previously seen in the B-band
(Mason et~al. 1980) and also present in the U and B light curves of
Fig.~\ref{fig3} to the end of the blue orbital hump centred at phase 
+0.25.

The eclipse varies from a round-bottomed, U-shape in the U-band
(suggesting a steep brightness distribution concentrated towards the 
disc centre) to a V-shape in the I-band (revealing a smoother 
and more extended brightness distribution). This is consistent with
the existence of a temperature gradient in the light source around 
the primary star, with the temperatures decreasing outwards.
This could arise from the expected radial temperature gradient in
the accretion disc, for an optically thin ADC, or from a vertically
stratified and decreasing temperature structure in an optically thick 
ADC (e.g., Iara et~al. 2001; Heinz \& Nowak 2001) veiling the inner 
accretion disc.

The end of the dip is seen as a positive jump in the light curves
around phase +0.13. The dip egress was measured by finding the phase of
maximum in the derivative of the light curve at the proper phase range
(cf. Wood, Irwin \& Pringle 1985).
The ingress of the dip in the I-band is seen as a sharp break in the
slope of the light curve and was similarly measured from its derivative. 
For the other bands, the starting phase of the dip was measured 
by finding the intersection between straight lines fitted to the data 
in the phase ranges prior (+0.4,+0.6) and after (+0.7,+0.8) the ingress 
of the dip and by finding the intersection of the line fitted to the 
latter phase range with the mean and median fluxes computed in the former
phase range. The results obtained with these procedures are consistent with
each other, with standard deviations smaller then 0.01~cycle in all bands.
Vertical ticks mark the measured ingress and egress phases of the dip 
in each light curve of Fig.~\ref{fig3}.

The starting phase of the dip is clearly wavelength-dependent, 
occurring {\em earlier} for shorter wavelengths 
(+0.61 in the U-band against +0.76 in the I-band). The observed 
behaviour in X-rays is consistent with this trend, with the X-ray dip
beginning as early as phase +0.4 (e.g., Heinz \& Nowak 2001).
This result also reflects the existence of a temperature gradient
in the disc or ADC and reveals that the thickening of the outer, 
occulting disc rim is gradual with azimuth at ingress.
On the other hand, the egress of the dip occurs at about the same phase
at all wavelengths, indicating that the thickness of the disc rim
decreases sharply at this azimuth. 
This is reminiscent of the azimuthal distribution of the outer 
accretion disc thickness found for the dwarf nova OY~Car during 
a superoutburst (Billington et~al. 1996), in which the disc rim
seems to increase abruptly close to the azimuthal position of the
bright spot and decreases slowly along the direction of rotation
of the gas in the disc, extending in azimuth for $\simeq 180\degr$.

We now turn our attention to the flickering curves.
The flickering curves have rather low signal-to-noise ratios since
they were derived from an ensemble of only 3-5 light curves.
Nevertheless, the analysis of the orbital flickering curves allow us to
constrain the location of the optical flickering sources in X1822-371.
If the flickering is produced in the bright spot where the gas stream
from the secondary star hits the outer edge of the disc (as it seems 
to occur in quiescent dwarf novae; e.g., Warner 1995, Bruch 2000),
the flickering curves should show an orbital hump (around phase +0.8) 
similar to that seen in the light curves of quiescent dwarf novae as well
as an asymmetric eclipse coincident in phase with the occultation of 
the bright spot by the secondary star (i.e., displaced towards positive
phases, see p.ex., Wood et~al. 1986; Baptista et~al. 1998).
On the other hand, if the flickering originates in the inner parts of 
the accretion disc (as is possibly the case in nova-like variables; 
see, for example, Horne \& Stiening 1985; Bruch 2000), then the 
spectrum of the flickering should be as blue as the spectrum of the 
inner disc regions and there should be an eclipse in the amplitude of 
the flickering following the occultation of the inner disc regions 
during the dipping phases. Such an eclipse should be more pronounced 
at shorter wavelengths.

Our flickering curves show a broad eclipse coincident with the dipping 
phases. The eclipse is less pronounced at longer wavelengths and
essentially disappears in the I-band. Whereas the eclipse is deep at
short wavelengths (1.4 mag in the B-band), it is not total.
Also, there is no sign of an orbital hump around phase +0.8 and no
evidence that the eclipse is displaced towards positive phases.
One can, therefore, discard the bright spot as an important source 
of flickering in X1822-371. On the other hand, the partial eclipse
and the changes in eclipse depth with wavelength suggest that the
flickering is associated to a multi-temperature accretion disc, 
with a `blue' component arising from the hot, inner disc regions 
(occulted during the dipping phases) and a `red' component probably
originating in the outermost disc regions (which remain visible at
mid-eclipse).
A more quantitative analysis of these results will be presented in 
section~\ref{fluxos}.

We computed the ratio between the flickering curve and the corresponding 
orbital light curve to derive the fractional contribution of the 
flickering component as a function of phase. 
We find that outside of the eclipse the flickering contributes 
$10\pm 2$ per cent of the total light, independent of wavelength.
During eclipse the fractional flickering contribution raises from 7 
per cent in the U- and B-bands to about 20 per cent in the I-band.
These values are significantly larger than the $\sim 2$ per cent
fractional flickering contribution estimated by Mason et~al. (1980)
from their unfiltered light curves. 
Their observations were made while the star was at $B\simeq 15.4$ mag, 
while our data correspond to a higher brightness state, $B\simeq 15.0$ mag.
This comparison indicates that the flickering activity in X1822-371 
might be correlated with the brightness level.
Further long-term, high-speed photometry would be useful to test this
hypothesis.

\subsection{The colors of the disc and of the flickering} \label{fluxos}

\begin{figure}
\includegraphics[bb=1.5cm 2cm 15cm 24cm,scale=0.51]{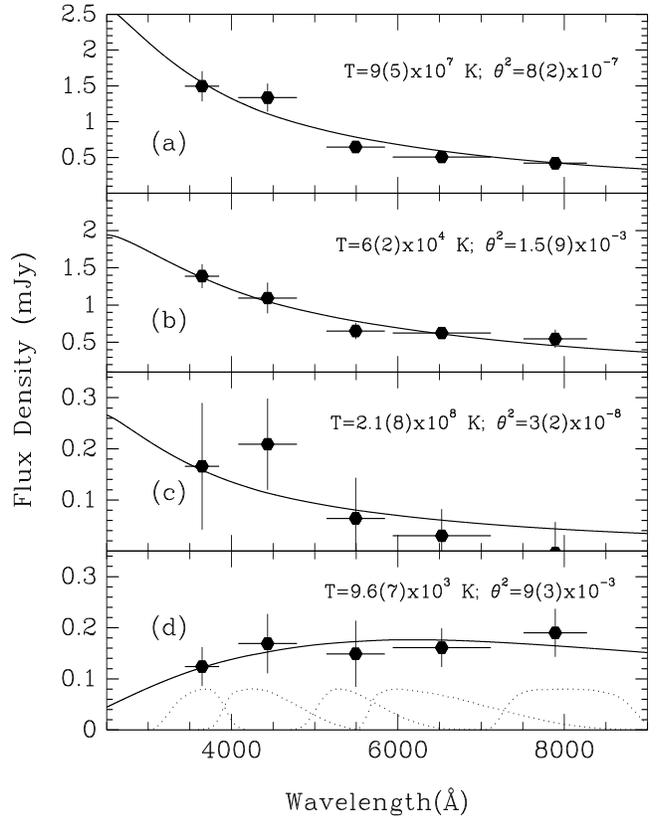}
 \caption{ Median fluxes for the different sources (filled symbols)
  with error bars (vertical ticks) and best fit blackbody models (solid 
  lines). Horizontal bars mark the full-width half-maximum of each 
  passband. (a) the inner disc regions, (b) the outer disc rim, (c) the 
  eclipsed flickering spectrum $(1-3)$, and (d) the uneclipsed flickering
  spectrum (region 1).
  The effective temperature, $T_{\rm eff}$, and solid angle, $\theta^2$, 
  of the best fit model are indicated in each panel. Dotted curves in 
  the lower panel show the normalized response function of the UBVRI 
  passbands. } 
\label{fig4}
\end{figure}

The UBVRI orbital light curves yield a unique opportunity to separate
the optical colors of the different light sources in the binary. 
For this purpose, we defined three phase ranges to extract median UBVRI 
fluxes, indicated by vertical dotted lines in Fig.~\ref{fig3}.
Region 1 (between phase +0.45 and +0.55) yields the colors of X1822-371
outside of eclipse and of the dip, where both the inner and the outer,
thick parts of the disc are visible. Region 2 (from phase $-0.2$ to $-0.1$)
corresponds to the phases where the outer, thick disc rim occults the
inner disc regions, while region 3 (from phase $-0.05$ to +0.05) corresponds
to the eclipse of the thick disc rim by the secondary star. Median UBVRI
fluxes were extracted from the average light curves for each region.
The difference between the fluxes of regions 1 and 2 yields the colors 
of the inner parts of the disc occulted during the dipping phases,
while the difference between the fluxes of regions 2 and 3 gives the
colors of the outer and thick disc rim covered by the secondary star.
The resulting fluxes are shown as filled circles in the two upper panels 
of Fig.~\ref{fig4}.
%
%
The slopes of the flux distributions are consistent with those of
optically thick thermal radiators.

A simple estimate of the dimensions and effective temperatures of these
light sources was obtained by fitting blackbody spectra to the observed 
fluxes using the {\sc iraf} synthetic photometry package SYNPHOT. 
Each fit assumes an interstellar extinction of $E(B-V)= 0.1$ (Mason 
\& Cordova 1982b).
The best-fit models are shown as solid lines in Fig.~\ref{fig4}. 
The corresponding effective temperature, $T_{\rm eff}$, and solid angle,
$\theta^2= \pi [(R/R_\odot)/(D/kpc)]^2$, are indicated in each panel. 
The uncertainties in the values of $T_{\rm eff}$ and $\theta^2$ were
obtained from Monte Carlo simulations by independently varying the 
values of the input UBVRI fluxes according to Gaussian distributions
with standard deviations equal to the corresponding uncertainties.

The temperature inferred for the inner disc regions is $T_{\rm eff}(12) 
= (9\pm 5) \times 10^7\;K$ for a solid angle of $\theta^2_{12}= 
(8\pm 2) \times 10^{-7}$.
A temperature of $T_{\rm eff}(23)= (6 \pm 2)\times 10^4\;K$ and a solid 
angle of $\theta^2_{23}= (1.5\pm 0.9) \times 10^{-3}$ were obtained for
the outer disc rim. 
The uncertainty in the fit of the inner disc regions is dominated 
by the error in the U-band flux and is large, since the slope of the 
blackbody spectrum at optical wavelengths is not sensitive to the 
temperature at these high temperatures.
We consistently find that the inner parts of the disc occulted during 
the dip are considerably hotter and smaller than the occulting outer 
disc rim. 
Since these sources are at the same distance, the ratio between the
fitted solid angles can be used to compare the characteristic 
dimension of the inner ($R_{12}$) and the outer ($R_{23}$) parts of the 
disc. We find $R_{12}= 0.023\;R_{23}$.

The colors of the optical flickering can be derived with a procedure
analogous to that used to separate the colors of the steady light sources.
Because in this case the data is of quite low signal-to-noise ratio, 
we adopted a simpler approach and computed the colors of the eclipsed 
flickering (the difference between the mean flickering amplitudes of 
regions 1 and 3) and of the flickering component that remains visible 
at mid-eclipse (the mean flickering amplitude of region 3), hereafter 
called `the uneclipsed flickering'.
The results are shown in the two lower panels of Fig.~\ref{fig4}.

As suggested in section~\ref{curvas}, the spectrum of the eclipsed
flickering component is very blue. A detailed comparison shows it is 
even bluer than the mean spectrum of the inner disc regions occulted 
during the dipping phases, signalling that the relative amplitude
of the eclipsed flickering increases towards shorter wavelengths.
Blackbody fits to the flickering colors yield temperatures of 
$T_{\rm eff}(eclip)= (2.1 \pm 0.8) \times 10^8\;K$ and of
$T_{\rm eff}(unecl)= (9.6\pm 0.7)\times 10^3\;K$, respectively,
for the eclipsed and the uneclipsed flickering components.
Since in this case we are fitting the spectrum of the fluctuations 
with respect to the mean UBVRI flux levels, the solid angles of these 
fits are meaningless.
Once again, the uncertainty in the temperature of the eclipsed flickering
is large because the slope of the blackbody spectrum at optical 
wavelengths is not sensitive to the temperature at these high temperatures.

\subsection{Time-series analysis} \label{fft}

In this section we further investigate the dependency in phase, as well
as in frequency, of the optical flickering in X1822-371. For this purpose, 
we computed Fast Fourier Transforms (e.g., Press et~al. 1992) of fragments 
of the individual UBVRI light curves centred at three selected phase 
ranges, one covering the eclipse (from phase $-0.1$ to +0.1), one 
sampling the dipping phases (from phase $-0.3$ to $-0.1$), and another 
sampling the regions outside of the eclipse and of the dip (from phase 
$-0.5$ to $-0.3$). In order to suppress the contribution from the
low-frequency components due to the eclipse, dip and orbital hump,
the corresponding average light curve (Fig.~\ref{fig3}) was 
subtracted from each fragment before computing the power spectrum.
The power spectra of the fragments of the individual light curves were
combined to compute median power spectra for each band and phase range.
Error bars were derived from the median of the absolute deviations
with respect to the median in each phase bin.

\begin{figure}
\scalebox{.6}{%
\includegraphics[bb=3cm 3.5cm 16cm 22cm,scale=1]{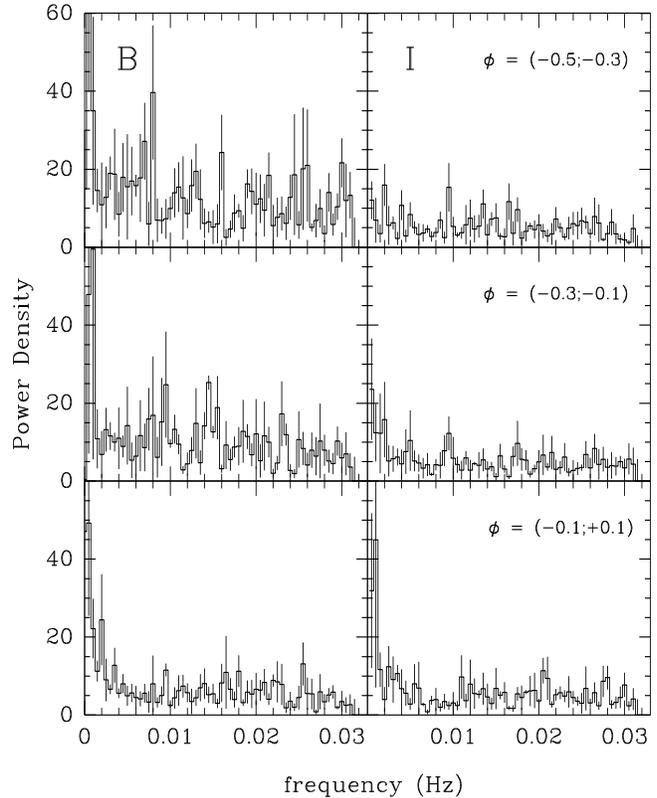} }%
 \caption{ Average power density spectra of selected portions of the 
  light curves in the B- and I-bands, at a resolution of $5 \times 
  10^{-4}\;Hz$. The uncertainties are shown as vertical ticks.
  The corresponding orbital phase range is indicated in each panel. } 
\label{fig5}
\end{figure}

Figure~\ref{fig5} shows average power spectra in the B- and I-bands 
for the three selected phase ranges.
%
%
The power spectra are flat, with no clear cut in frequency.
The I-band power spectra are consistent with each other at the 1-$\sigma$
level, confirming that there is no eclipse of the flickering component
at this wavelength. Similar results are found for the R-band.
On the other hand, the mean B-band power density is reduced by a factor
of $2.5 \pm 0.5$ during eclipse, in agreement with the results 
of section~\ref{curvas}. The behaviour is consistent at all frequencies
in the observed range, indicating that the blue, eclipsed flickering 
consists equally of low and high frequency components. 
The results are similar for the U-band, whereas the V-band is an 
intermediate case between those of the shorter and longer wavelengths.

\section{Conclusions} \label{fim}

The results of the analysis of high-speed optical photometry of the 
eclipsing X-ray binary X1822-371 can be summarized as follows:

\begin{enumerate}

\item We used new eclipse timings to derive a revised optical ephemeris. 
A quadratic fit to the eclipse timings is not statistically significant 
yet but suggests that the orbital period is increasing on a
timescale of $P_0/|\hbox{\.{P}}|= (4.2\pm 1.4) \times 10^6$~yr, in 
agreement with the results from the analysis of X-ray timings.
We also show that there is no systematic delay or advance of the optical
timings with respect to the X-ray timings.

\item The average UBVRI light curves show the deep eclipse of the disc
by the secondary star superimposed on the broader and shallower occultation
of the inner disc regions by the outer disc (dip), and an orbital hump
centred at phase +0.25 mostly seen in the U- and B-bands. 
There is no evidence of a secondary eclipse at phase +0.5 or of an 
ellipsoidal modulation of the secondary star light. 
The changes in eclipse shape with wavelength are consistent with the 
existence of a temperature gradient in the accretion disc and/or ADC.

\item The starting phase of the dip occurs {\em earlier} for shorter
wavelengths, while the egress occurs at the same phase in all bands. 
This suggests that the thickening of the outer, occulting disc rim is 
gradual with azimuth at ingress but decreases sharply at egress.

\item We used the dip and the eclipse to separate the colors of the
inner (occulted during the dip) and outer (eclipsed by the secondary star)
disc regions. The derived effective temperatures of the inner and outer 
disc regions are, respectively, $T_{\rm eff}= (9\pm 5)\times 10^7\;K$ 
and $T_{\rm eff}= (6\pm 2)\times 10^4\;K$, where we assumed an 
insterstellar extinction of $E(B-V)= 0.1$ for X1822-371.

\item The derived flickering curves show a broad eclipse coincident 
with the dipping phases, the depth of which decreases with increasing
wavelength. The blue, eclipsed flickering component is associated with 
the inner disc regions and can be fitted by a blackbody spectrum of 
$T_{\rm eff}= (2.1\pm 0.8) \times 10^8\;K$, whereas the uneclipsed
flickering component probably arises from the outermost disc regions
and is well described by a cooler blackbody of $T_{\rm eff}= 
(9.6\pm 0.7) \times 10^3\;K$. The power spectrum of the eclipsed 
flickering component is flat, with similar contributions from low
and high frequencies.

\item We find that outside of the eclipse the flickering contributes 
$10\pm 2$ per cent of the total light, independent of wavelength.
This is considerably larger than the 2 per cent fractional contribution
estimate of Mason et~al. (1980) and suggests that the flickering activity 
in X1822-371 is correlated with its brightness level.

\end{enumerate}

\section*{Acknowledgments}

We thank an anonymous referee for valuable comments and
suggestions on an earlier version of the manuscript.
This work was partially supported by the research grant FAURGS/PRONEX
7697.1003.00. RB acknowledges financial support from CNPq/Brazil through
grants 300\,354/96-7 and 520\,869/97-4.
AB acknowledges financial support from CAPES/Brazil and CNPq/Brazil.
ETH was supported by the TMR contract ERBFMBICT960971 of the European
Union.

\end{document}